\documentclass[aps,pre,twocolumn,groupedaddress,showpacs,preprintnumbers,amsmath]{revtex4-1}
\usepackage{epsfig,amsmath,amssymb,graphics,color,calc,hyperref}
\usepackage[utf8]{inputenc}
\usepackage{braket}
\usepackage{mathtools}
\usepackage{textcomp}
\newcommand{\sbrl}{\lfloor \! \lfloor}
\newcommand{\sbrr}{\rceil \! \rceil}

\begin{document}

\title{Mean-field phase diagram and spin glass phase of \\
the dipolar Kagome Ising antiferromagnet}

\author{Leticia F. Cugliandolo}
\affiliation{Sorbonne Universit\'{e}, Laboratoire de Physique Th\'{e}orique et Hautes Energies, CNRS
UMR 7589, 4, Place Jussieu, Tour 13, 5\`{e}me étage, 75252 Paris Cedex 05, France.}
\author{Laura Foini}
\affiliation{Institut de Physique Th\'{e}eorique, Universit\'{e} Paris Saclay, CNRS, CEA, F-91191
Gif-sur-Yvette, France}
\author{Marco Tarzia}
\affiliation{Sorbonne Universit\'{e}, Laboratoire de Physique Th\'eorique de la Mati\`ere Condens\'ee, CNRS UMR 7600, 4, Place Jussieu, Tour 13, 5\`{e}me étage, 75252 Paris Cedex 05, France}

\begin{abstract}
We derive the equilibrium phase diagram of the classical dipolar Ising antiferromagnet at the mean-field level on a geometry that mimics the two dimensional Kagome lattice. Our mean-field treatment is based on the combination of the cluster variational Bethe-Peierls formalism and the cavity method, developed in the context of the glass transition, and is
complementary to the Monte Carlo simulations realized in [Phys. Rev. B {\bf 98}, 144439 (2018)]. 
Our results confirm the nature of the low temperature crystalline phase which is reached through a weakly first-order phase transition.
Moreover, they allow us to interpret the dynamical slowing down observed in the work of Hamp \& al. as a remnant of a spin glass transition taking place at the mean-field level (and expected to be avoided in 2 dimensions).
\end{abstract}

\pacs{xxx}

\maketitle

\section{Introduction}

Many interesting classes of classical and quantum magnetic systems are extremely constrained. Hard local constraints lead to frustration and to the impossibility of satisfying all competing interactions simultaneously~\cite{Toulouse}, giving rise to the existence of highly degenerate ground states~\cite{Balents2010,Moessner2006}. Under certain conditions, these features produce a rich variety of collective behaviors~\cite{Balents2010,Moessner2006}, unconventional phase transitions~\cite{Jaubert2008,Lieb1972}, the emergence of a Coulomb phase with long-range correlations~\cite{Youngblood1980,Henley2010}, and other remarkably unusual and exotic phenomena.

On the other hand, frustration is also one of the key properties of glassy systems~\cite{Wolynes2012,Berthier2011,Cavagna2009,Gotze1991,Debenedetti2006,Tarjus2011}, where it also arises from the fact that minimizing some local interactions leads to the impossibility of minimizing other ones~\cite{Toulouse}. This feature can generate rugged energy landscapes and slow dynamics even in the absence of disorder~\cite{Marinari-etal94,BouchaudMezard,Cugliandolo-etal95,Bouchaud96,krzakala2008,Biroli2002,Ciamarra2003,Weigt2003,Rivoire2004,Tarzia2007,Tarjus2005,Westfahl2001,Grousson,Kurchan2012,Ritort2003,Chandler2010,Garrahan2000,plaquette}.

It is therefore surprising at first sight that very little is known on glassy phases in geometrically frustrated magnetic systems.
One of the first tentative investigations on this subject has been performed in Ref.~\cite{Sethna1992},
where glassy behavior was observed in nonrandomly frustrated Ising models with competing interactions.
More recently, strong nonequilibrium effects, slow dynamics, and super-Arrhenius relaxation have also been reported in two-dimensional spin systems with competing long-range and short-range interactions~\cite{Cannas2,Cannas}.

On a different front, a thermodynamic theory, called the ``frustration-limited domain theory''  of the properties of supercooled liquids, and of the extraordinary increase of their characteristic structural relaxation times as the temperature is lowered, was formulated in terms of the postulated existence of a narrowly avoided thermodynamic phase transition due to geometric frustration~\cite{Kivelson1995}
(see Ref.~\cite{Tarjus2005} for a review). In this context frustration describes an incompatibility between extension of the locally preferred order in a liquid and tiling of the whole space. 
This picture is consistent with appropriate minimal statistical mechanical models, such as three-dimensional Ising Coulomb frustrated lattice models, which display a slowing down of the relaxation in Monte Carlo simulations~\cite{Grousson,TarjusAvoided} and an ideal glass transition within mean-field approximations~\cite{Westfahl2001,Grousson}.
However, numerical simulations of these models in $3d$ are limited by the presence of a first-order transition to a modulated, defect-ordered phase~\cite{Grousson}, and cannot be performed at sufficiently low temperatures.

Several frustrated spin (or Potts) lattice models without quenched disorder have also been introduced and studied over the past years to describe the key features of the glass transition. However, most of them are either mean-field in nature (and cannot be easily generalized to finite dimensions)~\cite{Marinari-etal94,BouchaudMezard,Cugliandolo-etal95},  
or are characterized by (unphysical) multi-body interactions~\cite{plaquette}.
The classical three-coloring model on the two-dimensional hexagonal lattice have been shown to undergo a dynamical freezing in metastable states very similar to the one observed in structural glasses~\cite{cepas2012,cepas2014}.
Slow dynamics also appears in electronic Coulomb liquids on the triangular lattice at quarter-filling~\cite{Dobro2015}, as well as in spin-ice systems both in $2d$~\cite{Budrikis2012,Levis2012} and in $3d$~\cite{Harris1997}.
On the quantum side, it was shown in Ref.~\cite{VBG} that a valence bond glass phase emerges in the SU(N) Hubbard-Heisenberg model on a Bethe lattice in the large-$N$ limit due to the interplay of strong magnetic frustration and quantum fluctuations. 

Yet, despite all these efforts over the past years, a clear and coherent picture of the glassy behavior that can arise due to the effect of geometric frustration in finite dimensional magnetic systems at low temperature is still missing.

Recently Hamp \& al.~\cite{Hamp2018} studied an Ising model on the Kagome lattice with short range antiferromagnetic interactions and dipolar interactions decaying as $1/r^3$---the Dipolar Kagome Ising Antiferromagnet (DKIAFM) introduced in~\cite{Chioar2016}. 
By means of extensive Monte Carlo simulations the authors first showed evidence for a first-order transition from the high temperature paramagnetic phase to a low temperature  crystal state that breaks time-reversal and sublattice symmetries, and coincides with the one previously proposed in Ref.~\cite{Chioar2016} as the ground state.
Furthermore, upon cooling below the first-order transition, the system enters a supercooled liquid regime which exhibits all the characteristic features of fragile glasses: 
two-time autocorrelation functions decay as stretched exponentials and the relaxation time grows in a super-Arrhenius fashion as the temperature is decreased.
However, these conclusions were drawn out of numerical simulations of relatively small systems (about $300$ spins) and might be affected by both strong finite-size effects and the difficulty of reaching thermal equilibrium in a reliable fashion due to strong metastability effects.
Moreover, a consistent picture of the physical origin of the dynamical slowing down at low temperatures has not been convincingly established yet.

In order to overcome, at least partially, these issues, in this paper we perform an analytical study 
of the equilibrium phase diagram of the DKIAFM in the thermodynamic limit at a mean-field level, focusing both on the ordered state and the glassy phase. Our results essentially confirm, support, and elucidate the observations reported in Ref.~\cite{Hamp2018}.
Upon decreasing temperature, we first find a transition to a six-fold degenerate crystal state which breaks time reversal and rotation sublattice symmetry as the one observed in~\cite{Hamp2018,Chioar2016}.
The mean-field analysis indicates that the transition is indeed discontinuous. However, its first-order nature turns out to be extremely weak: the spinodal point of the crystal phase is very close to the transition point, resulting in a very large jump of the specific heat at the transition.  This feature provides a possible explanation of the fact that the finite-size scaling of the numerical data of the maximum of the specific heat with the system size performed in Ref.~\cite{Hamp2018} did not find the usual behavior ($C_{\rm max} \propto N$) expected at a first-order transition due to very large finite-size effects.

When the system is supercooled below the first-order transition, we find that the paramagnetic state becomes unstable below a temperature at which the spin glass susceptibility diverges. Here, a continuous spin glass transition takes place at the mean-field level~\cite{Mezard1987}.

Note that the fact that the model displays a continuous spin glass transition in mean-field instead of a Random First-Order Transition~\cite{Kirkpatrick1989,Lubchenko2007} 
of the kind found in structural glasses (such as hard spheres in infinite dimensions~\cite{Kurchan2012} and lattice glass models on the Bethe lattice~\cite{Biroli2002,Ciamarra2003,Rivoire2004,Tarzia2007,krzakala2008}) is perhaps not surprising. In fact, Ising spins with antiferromagnetic couplings on high-dimensional frustrated lattices and other related frustrated mean-field models with pairwise interactions are known to undergo a continuous transition to a spin glass phase 
when the temperature is lowered below the critical temperature of the antiferromagnetic phase~\cite{Mezard2001,krzakala2008}.

Beyond the fact that both spin glasses and structural glasses exhibit a pronounced slowing down of the dynamics upon cooling and aging in the low temperature phase, several important qualitative and quantitative differences characterize the dynamical behavior of these systems: In structural glasses two-time autocorrelation functions generically exhibit a two-step relaxation, characterized by a relatively fast decay to a plateau (i.e., the Edwards-Anderson order parameter) which appears discontinuously upon lowering the temperature, followed by a much slower decay, described by a stretched exponential. Moreover, the structural relaxation time is found to grow extremely fast, in a super-Arrhenius fashion, as the temperature is decreased~\cite{Wolynes2012,Berthier2011,Cavagna2009,Gotze1991,Debenedetti2006,Tarjus2011}. Conversely, in spin glasses the Edwards-Anderson order parameter is continuous at the transition and vanishes in the paramagnetic phase. Hence two-time autocorrelation functions should display a simple exponential decay when the transition is approached from the high temperature phase, and an algebraic decay at the critical point. Furthemore, the relaxation time is expected to diverge (only) as a power-law at the critical point (and to stay infinite in the whole low temperature phase)~\cite{Mezard1987}.
Nonetheless these differences are not clearly visible in numerical simulation of relatively small samples. A clear example of that is provided by the analysis of the dynamics of $3d$ Ising spin glasses performed in Ref.~\cite{Ogielski1985,OgielskiMorgenstern1985} using Monte Carlo simulations of systems with up to $64^3$ spins. The two-time autocorrelation function was found to be very well fitted by stretched exponentials, with an exponent $\beta$ that exhibits a temperature dependence extremely similar to the one reported in~\cite{Hamp2018} for the DKIAFM. Moreover, although the divergence of the relaxation time as a power law, $\tau \sim (T - T_c)^{-z \nu}$, is consistent with the numerics, a Vogel-Fulcher law, $\tau \sim e^{E_0/(T-T_c)}$, was also found to account reasonably well for the data.

The lower-critical dimension of the spin glass transition is expected to be $d_L \approx 2.5$~\cite{Franz1994} (at least in the case of short-range interactions). Hence on general grounds we do not expect a genuine spin glass phase for the DKIAFM in $2d$.
Yet, the manifestations of the vestige of the transition can be very strong also in two dimensional systems: The spin glass amorphous order can establish over very long (although not infinite) length scales, the spin glass susceptibility can become very large (although not infinite), and the relaxation time can grow very fast at low temperature.
Several experimental realizations of two-dimensional spin glasses using thin films do indeed show the same behavior as $3d$ spin glasses at sufficiently low temperature~\cite{Mattson1992,Gucchhait2017,Fernandez2019}.
In this sense, the existence of the spin glass phase in higher dimension, accompanied by the growth of long-range amorphous order and a rough free-energy landscape, provides a possible and natural explanation of the slow dynamics observed in the numerical simulations of Ref.~\cite{Hamp2018} of the $2d$ model at low temperatures.

Despite the fact that our mean-field approach consists in studying the model on a random sparse graph of triangular Kagome plaquettes, and cutting-off the dipolar interactions beyond the second nearest-neighbour plaquettes (i.e., the $5$th nearest-neighbor spins), it provides a remarkably good approximation for the equilibrium properties of the $2d$ DKIAFM. For instance, the mean-field approach yields a zero temperature entropy density of the nearest-neighbor Kagome spin ice model obtained for $D=0$~\cite{Wills2002} equal to $s_{\rm GS} \approx 0.75204$, which turns out to be extremely close to the Pauling estimate $s_{\rm GS} \approx 0.75225$. Similarly, the ground state energy density of the crystal ground state within the mean-field approximation is $e_{\rm GS} \approx -1.6116$, which accounts reasonably well for the one found in the Monte Carlo simulations of systems with $300$ spins, $e_{\rm GS} \approx -1.515$~\cite{remark}. As expected, the transition temperature to the crystalline phase is overestimated (by about a factor $3$) by the mean-field treatment. Yet, the temperature dependence of the specific heat, the energy density, and the magnetization are remarkably similar, also at a quantitative level, to the ones found with Monte Carlo simulations (see Fig.~\ref{fig:observables}).

The results presented here can serve at least two purposes:
(i) They help to support, understand, and clarify the numerical results of Ref.~\cite{Hamp2018}.
(ii) They provide a first step to bridge the gap between the slow dynamics observed in geometrically frustrated magnetic systems and the theory of the glass transition formulated in terms of rough free-energy landscapes.
We believe that this analysis is of particular interest, especially in the light of the experimental relevance of the model, which could be potentially realized in several realistic set ups, including colloidal crystals~\cite{Han2008,Zhou2017}, artificial nanomagnetic arrays~\cite{Chioar2014,Nisoli2013}, cold polar molecules~\cite{Ni2008},  atomic gases with large magnetic dipole moments~\cite{Griesmaier2005}, and layered bulk Kagome materials~\cite{Scheie2016,Paddison2016,Dun2017}.

The paper is organized as follows. In the next section we introduce the model. In Sec.~\ref{sec:mf} we describe the mean-field approach, based on a cluster formulation of the problem on the Bethe lattice. In Sec.~\ref{sec:results} we show the results found within our analytical treatment, including the phase diagram 
and the equation of state. Finally, in Sec.~\ref{sec:conclusions} we provide some concluding remarks and perspectives for future work.

\section{The model}

We consider the DKIAFM~\cite{Hamp2018,Chioar2016}
in which $N$ classical spins $S_i = \pm 1$ are
placed on the vertices of a two-dimensional 
Kagome lattice and point in a direction
perpendicular to the plane.
The Hamiltonian comprises an antiferromagnetic exchange term of
strength $J$ between spins at nearest-neighbor lattice sites
$\langle ij \rangle$ and long-range dipolar interactions of characteristic
strength $D$ between all pairs of spins:
\begin{equation} \label{eq:H}
    {\cal H} = J \sum_{\langle ij \rangle} S_i S_j + \frac{D}{2} \sum_{i \neq j} \frac{S_i S_j}{r_{ij}^3} 
    \; ,
\end{equation}
where the distance $r_{ij} = | {\bf r}_i - {\bf r}_j|/a$ between the spins $i$ and $j$ is measured in units of the lattice spacing $a$ (that we set equal to $1$ throughout).

In the following we will be interested in the case in which both interactions are antiferromagnetic, i.e., $J>0$ and $D>0$.
The case $D = 0$ is known to be fully frustrated and
does not order down to zero temperature~\cite{Takagi1993}. 

The phase
diagram of the $J = 0$ model is less well understood
but the system is again strongly frustrated with any ordering (if present at all) suppressed down to temperatures
$T \ll D$~\cite{Chioar2014}.

The previous studies of the model~\cite{Hamp2018,Chioar2016} considered  the  coupling  parameters $D= 1 \, {\rm K}$ and $J = 0.5 \, {\rm K}$ (setting $k_B= 1$ and measuring all energies in Kelvin). A further advantage of developing an analytic (although approximate) treatment is that it is relatively simple to explore the parameter space. Without loss of generality we set $J=0.5 \, {\rm K}$ throughout (as in~\cite{Hamp2018,Chioar2016}) and study the phase diagram of the model and the constitutive equations in the different phases varying the dipolar coupling $D$ and the temperature $T$.

\begin{figure}
\includegraphics[width=0.46\textwidth]{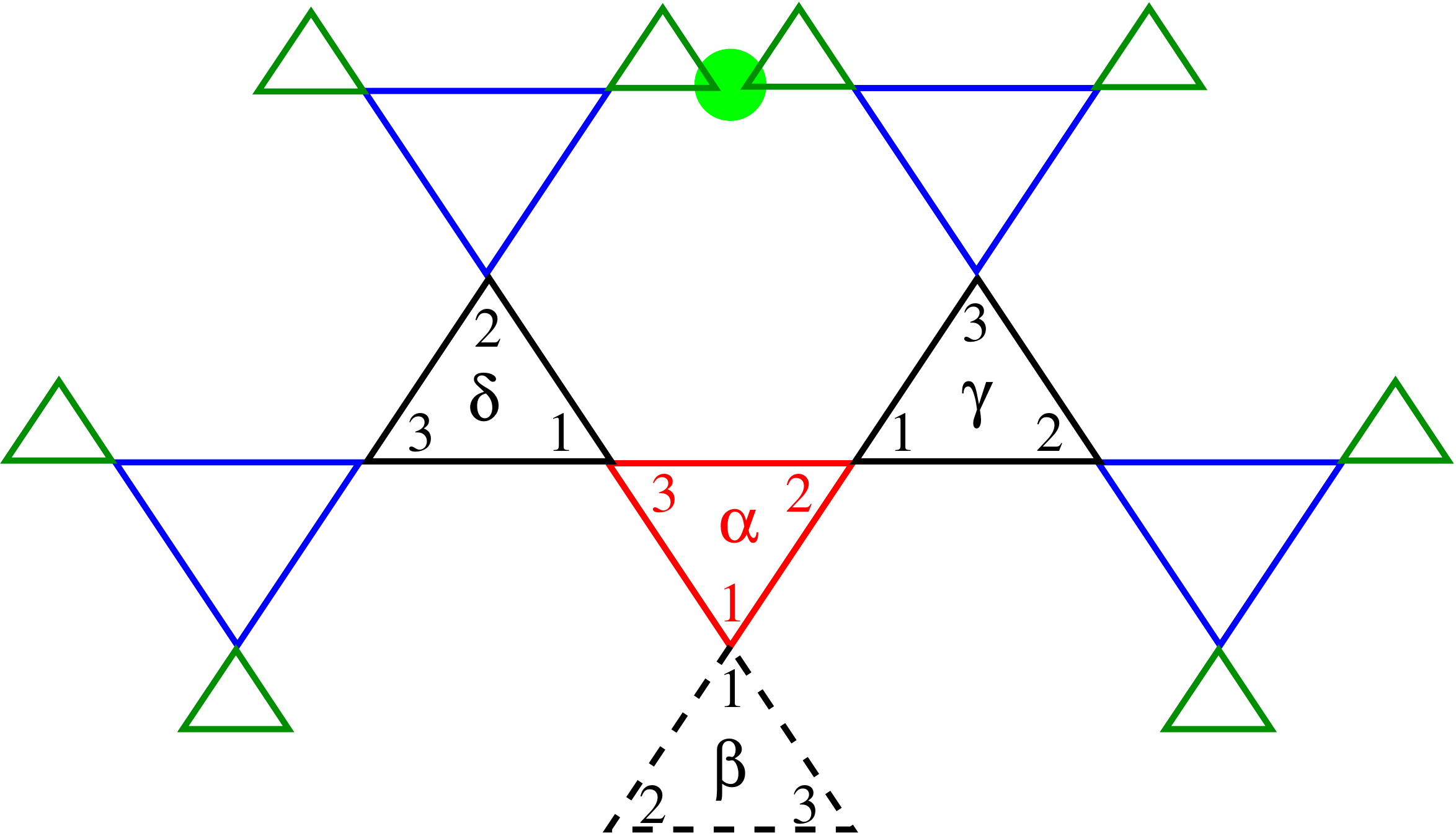}
\vspace{0.2cm}
\caption{\label{fig:sketch} Sketch of a small portion of a (rooted) Random Regular Graph (RRG) of triangular plaquettes in presence of a cavity (the dashed black plaquette $\beta$). Each up-type triangular plaquette of the RRG is connected to three down-type triangular plaquettes and each down-type triangular plaquette is connected to up down-type triangular plaquettes. The graph looks locally like a tree since typical loops are very large (the typical size of the loops diverges as $\log N_\bigtriangleup$). The first nearest-neighboring plaquettes ($\beta$, $\gamma$, and $\delta$) of the central (red) triangle ($\alpha$) are drawn in black, the second nearest-neigboring plaquettes in blue and the third nearest-neighboring plaquettes in green. The mean-field Bethe-Pierles approximation consists in discarding the fact that the two spins inside the green circle are in fact the same spin on the original kagome lattice. Moreover the dipolar coupling is cut-off beyond the second nearest-neighboring plaquettes (i.e., $5$th nearest-neighbor spins).}
\end{figure}

\section{The mean-field analysis} \label{sec:mf}

Our mean-field treatment is based on the combination of the cluster variational Bethe-Peierls
formalism (already successfully employed in the study of the equilibrium properties of geometrically frustrated magnetic systems~\cite{Jaubert2008,Pelizzola2005,Levis2013,Foini2013,Cugliandolo2015}) and the cavity method~\cite{Mezard2001}, developed in the context of glassy and disordered systems described by replica symmetry breaking (RSB). The latter concept is related to a complex free-energy landscape with special structure and the calculational meaning of it, in the context of the cavity method, will become clear below. 
In particular, we define the model on a Random Regular Graph~\cite{RRG} (RRG) of 
$N_\bigtriangleup$ triangular plaquettes of
total coordination three (see Fig.~\ref{fig:sketch} for a sketch). In this case, the number of spins is equal to $N=3 N_\bigtriangleup/2$ since since each spin belongs to two plaquettes and each plaquette contains three spins.
RRGs are a special class of sparse graphs, whose elements are chosen at random with uniform probability over the ensembles of all graphs of $N_\bigtriangleup$ nodes, such that each node (i.e., a triangular plaquette) has exactly three neighbors.
RRGs have a local tree-like structure, which allows one to obtain exact self-consistent recursion relations for the probability
distributions of the spin configurations on each plaquette of the graph.
Yet, they have large loops, whose typical
length scales as $\log N_\bigtriangleup$ and diverges in the thermodynamic limit. Hence, a RRG 
is locally a tree, but it is frustrated, does not have a boundary, and is statistically translational invariant. For these reasons RRGs are
suitable lattices to study the thermodynamics of glassy and disordered systems at a mean-field level~\cite{Biroli2002,Ciamarra2003,Rivoire2004,Tarzia2007} (i.e., in the limit of infinite dimensions). 

\subsection{The ``cavity'' recursion relations} \label{sec:cavity}

The standard way to obtain the recursion relations for the  marginal  probabilities of observing a given spin configuration on a given plaquette is provided by the {\it cavity 
method}~\cite{Mezard2001}, which is equivalent to the Bethe-Peierls approximation at the replica-symmetric level. The cavity method is based on the assumption that, due to the tree-like structure of the lattice, in absence of a given plaquette (the {\it cavity}, e.g. the red triangle of Fig.~\ref{fig:sketch}), the neighboring plaquettes (the black triangles of Fig.~\ref{fig:sketch}) are uncorrelated and their marginal joint probabilities factorize.
Thanks to such factorization property one can write relatively simple recursion equations for the marginal probabilities of the cavity sites. Such equations have to be solved self-consistently, the fixed points of which yield the free-energy of the system along with all the thermodynamic observables (all the technical details of the method can be found in Refs.~\cite{Mezard2001,Rivoire2004}). However, in order to be tractable, the cavity approach is formulated for systems with finite range interactions. Hence, before proceeding further we need to treat the dipolar interactions of Eq.~(\ref{eq:H}) in an approximate fashion. In practice, in the analytic calculations described below we choose to cut-off the dipolar couplings up to second nearest-neighboring plaquettes (i.e., the interactions between the spins belonging to the red plaquette $\alpha$ of Fig.~\ref{fig:sketch} and the spins belonging to green plaquettes are set to zero).

Consider now the cavity triangle $\alpha$ (red) in absence of one of its neighboring plaquettes $\beta$ (dashed black). We define $p_{\alpha \to \beta} (\{S_\alpha\}|\{S_\beta\})$ as the probability to observe the spin configuration $\{S_\alpha\} \equiv \{ S_{\alpha,1},  S_{\alpha,2},  S_{\alpha,3} \}$ on the cavity triangle $\alpha$ of the (rooted) RRG, given that the spin configuration of the plaquette $\beta$ is $\{S_\beta\} \equiv \{ S_{\beta,1},  S_{\beta,2},  S_{\beta,3} \}$.
We have adopted the convention that spin $1$ is the root of the cavity plaquette and spins $2$ and $3$ are labeled anticlockwise. Using this convention one has that $ S_{\beta,1} \equiv S_{\alpha,1}$.

The probabilities $p_{\alpha \to \beta} (\{S_\alpha\}|\{S_\beta\})$ can be written in terms of the marginal probabilities defined on the cavity triangles $\gamma$ and $\delta$ in absence of the triangle $\alpha$, times the Gibbs' weight associated to each spin configuration:
\begin{widetext}
\begin{equation} \label{eq:cavity}
\begin{aligned}
    p_{\alpha \to \beta}(\{S_\alpha\}|\{S_\beta\})& = \left( {\cal Z}_{\alpha \to \beta}^{({\rm iter})} \right)^{-1} \!\! \sum_{ \sbrl \gamma, \delta \sbrr_\alpha}  
    \! p_{\gamma \to \alpha}(\{S_\gamma\}|\{S_\alpha\}) \, p_{\delta \to \alpha}(\{S_\delta\}|\{S_\alpha\}) 
    \, e^{-\beta \tilde{{\cal H}}_{\alpha \to \beta}  (\{S_\alpha 
    , 
    S_\gamma 
    , 
    S_\delta \}
    | \{
    S_\beta\}) } 
    \, ,
    \end{aligned}
\end{equation}
where ${\cal Z}_{\alpha \to \beta}^{({\rm iter})}$ is a normalization factor ensuring that $\sum_{\{S_\alpha\},\{S_\beta\}} p_{\alpha \to \beta}(\{S_\alpha\}|\{S_\beta\}) = 1$
and is associated to the ``free-energy shift'' involved in the iteration process:
$- \beta \Delta F_{\alpha \to \beta}^{({\rm iter})} \equiv \log {\cal Z}_{\alpha \to \beta}^{({\rm iter})}$.
Here we introduce the notation $\sum_{\sbrl \gamma, \delta \sbrr_\alpha}$ that indicates the sum over all possible configurations $\{S_\gamma \}$ and $\{S_\delta \}$ of the spin degrees of freedom of the plaquettes $\gamma$ and $\delta$, compatible with the constraints imposed by the spin configuration $\{S_\alpha \}$ on the plaquette $\alpha$, i.e., $S_{\gamma,1} = S_{\alpha,2}$ and $S_{\delta,1} = S_{\alpha,3}$ [see Eq.~(\ref{eq:sums})].
We will use this notation throughout this section.
The Hamiltonian 
$\tilde{{\cal H}}_{\alpha \to \beta} (\{S_\alpha 
,
S_\gamma
,
S_\delta\} |
\{
S_\beta\})$ appearing in the Gibbs factor of Eq.~(\ref{eq:cavity}) is a modified Hamiltonian, Eq.~(\ref{eq:H}), restricted to the cavity plaquette $\alpha \to \beta$:
\begin{equation} \label{eq:Hres}
\begin{aligned}
\tilde{{\cal H}}_{\alpha \to \beta} (\{S_\alpha ,S_\gamma,S_\delta\} | \{ S_\beta\})
& \equiv {\cal H}_\alpha^{({\rm AF})} + \tilde{{\cal H}}_{\gamma \to \alpha}^{(2{\rm D})} + \tilde{{\cal H}}_{\delta \to \alpha}^{(3{\rm D})} + \tilde{{\cal H}}_{(\gamma,\delta)}^{(2{\rm NND})} + \tilde{{\cal H}}_{(\gamma,\beta)}^{(2{\rm NND})} + \tilde{{\cal H}}_{(\delta,\beta)}^{(2{\rm NND})} \, , \\
{\cal H}_\alpha^{({\rm AF})} & = \left( J + D \right ) \left[ S_{\alpha,1} S_{\alpha,2} + S_{\alpha,1} S_{\alpha,3} + S_{\alpha,2} S_{\alpha,3} \right] \, , \\
\tilde{{\cal H}}_{\gamma \to \alpha}^{(2{\rm D})} & = D \left [ \frac{ S_{\alpha,1} S_{\gamma,2} + S_{\alpha,3} S_{\gamma,3} }{3 \sqrt{3}} + \frac{S_{\alpha,1} S_{\gamma,3} + S_{\alpha,3} S_{\gamma,2}}{8} \right] \, , \\
\tilde{{\cal H}}_{\delta \to \alpha}^{(3{\rm D})} & = D \left [ \frac{ S_{\alpha,1} S_{\delta,3} + S_{\alpha,2} S_{\delta,2} }{3 \sqrt{3}} + \frac{S_{\alpha,1} S_{\delta,2} + S_{\alpha,2} S_{\delta,3}}{8} \right] \, , \\
\tilde{{\cal H}}_{(\gamma,\delta)}^{(2{\rm NND})} & = D \left [ \frac{S_{\gamma,3} S_{\delta,2}}{8} + \frac{ S_{\gamma,2} S_{\delta,2} + S_{\gamma,3} S_{\delta,3} }{7 \sqrt{7}} + \frac{S_{\gamma,2} S_{\delta,3}}{27} \right ] \, .
\end{aligned}
\end{equation}
The meaning of this decomposition is the following. $\tilde{{\cal H}}_{\gamma \to \alpha}^{(2{\rm D})}$ contains the $4$ dipolar interaction terms between the spins belonging to the plaquette $\alpha$ and the spin belonging to its nearest-neighbor plaquette $\gamma$ attached to the spin $S_{\alpha,2}$, which are not already contained in ${\cal H}_\alpha^{({\rm AF})}$. $\tilde{{\cal H}}_{(\gamma,\delta)}^{(2{\rm NND})}$ contains the $4$ dipolar interaction terms between the spins of the second nearest-neighbor plaquettes $\gamma$ and $\delta$ which are not already contained in $\tilde{{\cal H}}_{\gamma \to \alpha}^{(2{\rm D})}$ and $\tilde{{\cal H}}_{\delta \to \alpha}^{(3{\rm D})}$.

In order to obtain the marginal probabilities of the spin configurations on each plaquette of the (unrooted) RRG (where each triangular plaquette has exactly three neighbors), one needs to merge three cavity plaquettes (e.g., plaquettes $\beta$, $\gamma$, and $\delta$ of Fig.~\ref{fig:sketch}) onto their neighboring plaquette (e.g., plaquette $\alpha$ of Fig.~\ref{fig:sketch}). In this way one obtains:
\begin{equation} \label{eq:marginals}
\begin{aligned}
    P_\alpha(\{S_\alpha\})& = \left( {\cal Z}_{\alpha}^{(s)} \right)^{-1} \!\!\! \sum_{
    \sbrl \beta, \gamma, \delta \sbrr_\alpha}
    \! p_{\beta \to \alpha}(\{S_\beta\}|\{S_\alpha\}) \, 
    p_{\gamma \to \alpha}(\{S_\gamma\}|\{S_\alpha\}) \, p_{\delta \to \alpha}(\{S_\delta\}|\{S_\alpha\}) 
    \, e^{-\beta \tilde{{\cal H}}_\alpha (\{S_\alpha
    ,
    S_\beta
    ,
    S_\gamma
    ,
    S_\delta\})} 
    \, ,
    \end{aligned}
\end{equation}
where ${\cal Z}_{\alpha}^{(s)}$ is a normalization factor ensuring that $\sum_{\{S_\alpha\}} P_\alpha(\{S_\alpha\}) = 1$, and is associated to the ``free-energy shift'' involved in the process of joining three cavity plaquettes ($\beta$, $\gamma$, and $\delta$) to a central plaquette ($\alpha$): $- \beta \Delta F_\alpha^{(s)} \equiv \log {\cal Z}_{\alpha}^{(s)}$.
The plaquette Hamiltonian $\tilde{{\cal H}}_\alpha (\{S_\alpha,S_\beta,S_\gamma,S_\delta\})$ reads
\begin{equation} \label{eq:Hres1}
\begin{aligned}
\tilde{{\cal H}}_{\alpha} (\{S_\alpha ,S_\beta, S_\gamma,S_\delta\})
& \equiv {\cal H}_\alpha^{({\rm AF})} + \tilde{{\cal H}}_{\beta \to \alpha}^{(1{\rm D})} +
\tilde{{\cal H}}_{\gamma \to \alpha}^{(2{\rm D})} + \tilde{{\cal H}}_{\delta \to \alpha}^{(3{\rm D})} + \tilde{{\cal H}}_{(\gamma,\delta)}^{(2{\rm NND})} + \tilde{{\cal H}}_{(\gamma,\beta)}^{(2{\rm NND})} + \tilde{{\cal H}}_{(\delta,\beta)}^{(2{\rm NND})} \, , \\
\tilde{{\cal H}}_{\beta \to \alpha}^{(1{\rm D})} & = D \left [ \frac{ S_{\alpha,2} S_{\beta,3} + S_{\alpha,3} S_{\beta,2} }{3 \sqrt{3}} + \frac{S_{\alpha,2} S_{\beta,2} + S_{\alpha,3} S_{\beta,3}}{8} \right] \, ,
\end{aligned}
\end{equation}
and the other terms are given in Eq.~(\ref{eq:Hres}).

The equilibrium averages of all local observables which involve the spin degrees of freedom of a given plaquette, including, e.g.,  the magnetization, can be expressed in terms of these marginal probabilities: 
\begin{equation} \label{eq:obs}
\langle O_\alpha \rangle = 
\sum_{\{S_\alpha \}} O (\{S_\alpha\}) P_\alpha(\{S_\alpha\}) \, .
\end{equation}
Similarly, the contribution to the average energy due to the plaquette $\alpha$ can be expressed as
\begin{equation} \label{eq:energy1}
\begin{aligned}
    \left \langle e_\alpha^{(s)} \right \rangle 
    & \! = \! \frac{\sum\limits_{ \sbrl \alpha \to (\beta,\gamma,\delta) \sbrr}  
    p_{\beta \to \alpha}(\{S_\beta\}|\{S_\alpha\}) \, 
    p_{\gamma \to \alpha}(\{S_\gamma\}|\{S_\alpha\}) \, p_{\delta \to \alpha}(\{S_\delta\}|\{S_\alpha\}) 
    \, \tilde{{\cal H}}_{\alpha} (\{S_\alpha,S_\beta,S_\gamma,S_\delta\}) \,
    e^{-\beta \tilde{{\cal H}}_\alpha (\{S_\alpha
    ,
    S_\beta
    ,
    S_\gamma
    ,
    S_\delta\})}}
    {\sum\limits_{ \sbrl \alpha \to ( \beta, \gamma, \delta ) \sbrr}  
    \! p_{\beta \to \alpha}(\{S_\beta\}|\{S_\alpha\}) \, 
    p_{\gamma \to \alpha}(\{S_\gamma\}|\{S_\alpha\}) \, p_{\delta \to \alpha}(\{S_\delta\}|\{S_\alpha\}) 
    \, e^{-\beta \tilde{{\cal H}}_\alpha (\{S_\alpha
    ,
    S_\beta
    ,
    S_\gamma
    ,
    S_\delta\})}}
    \, .
    \end{aligned}
\end{equation}
The process of joining two neighboring cavity plaquettes (e.g., plaquettes $\alpha$ and $\beta$ of Fig.~\ref{fig:sketch}) involves another ``free-energy shift'', defined as
\begin{equation} \label{eq:DFl}
e^{- \beta \Delta F_{\alpha \leftrightarrow \beta}^{(l)}} \equiv {\cal Z}_{\alpha \leftrightarrow \beta}^{(l)} = \sum_{\sbrl \alpha \leftrightarrow \beta \sbrr}
    p_{\alpha \to \beta}(\{S_\alpha\}|\{S_\beta\}) \, 
    p_{\beta \to \alpha}(\{S_\beta\}|\{S_\alpha\}) \, 
    e^{-\beta \tilde{{\cal H}}_{\beta \to \alpha}^{(1{\rm D})}} \, ,
\end{equation}
where the Gibbs' factor $\tilde{{\cal H}}_{\beta \to \alpha}^{({\rm D})}$ has been defined in Eq.~(\ref{eq:Hres1}). Similarly, the contribution to the average energy coming from the interactions between two neighboring cavity plaquettes is given by
\begin{equation} \label{eq:energy2}
\left \langle e_{\alpha \leftrightarrow \beta}^{(l)} \right \rangle = \frac{
\sum\limits_{\sbrl \alpha \leftrightarrow \beta \sbrr}
    p_{\alpha \to \beta}(\{S_\alpha\}|\{S_\beta\}) \, 
    p_{\beta \to \alpha}(\{S_\beta\}|\{S_\alpha\}) \, 
    \tilde{{\cal H}}_{\beta \to \alpha}^{(1{\rm D})} \,
    e^{-\beta \tilde{{\cal H}}_{\beta \to \alpha}^{(1{\rm D})}}}
    {\sum\limits_{\sbrl \alpha \leftrightarrow \beta \sbrr}
    p_{\alpha \to \beta}(\{S_\alpha\}|\{S_\beta\}) \, 
    p_{\beta \to \alpha}(\{S_\beta\}|\{S_\alpha\}) \, 
    e^{-\beta \tilde{{\cal H}}_{\beta \to \alpha}^{(1{\rm D})}}} \, .
\end{equation}
We recall here the convention adopted for the notation of the summation over the spin degrees of freedom in the expressions above:
\begin{equation} \label{eq:sums}
\begin{aligned}
\sum_{ \sbrl \gamma, \delta \sbrr_\alpha} &\equiv \sum_{ \substack{\{S_\gamma\},\{S_\delta\}\\  S_{\gamma,1} = S_{\alpha,2} \\ S_{\delta,1} = S_{\alpha,3}}} \, , \qquad \qquad \qquad \qquad  \qquad \qquad \qquad \qquad 
    \sum_{ \sbrl \beta, \gamma, \delta \sbrr_\alpha} \equiv \sum_{ \substack{\{S_\beta \}, \{S_\gamma\},\{S_\delta\}\\ 
    S_{\beta,1} = S_{\alpha,1} \\
    S_{\gamma,1} = S_{\alpha,2} \\ S_{\delta,1} = S_{\alpha,3}}} \, , \\ 
   \sum_{ \sbrl \alpha \to (\beta, \gamma, \delta) \sbrr} & \equiv \sum_{ \substack{\{S_\alpha \},\{S_\beta \}, \{S_\gamma\},\{S_\delta\}\\ 
    S_{\beta,1} = S_{\alpha,1} \\
    S_{\gamma,1} = S_{\alpha,2} \\ S_{\delta,1} = S_{\alpha,3}}} 
    = \sum_{\{S_\alpha \}} \sum_{ \sbrl \beta, \gamma, \delta \sbrr_\alpha}
    \, , \qquad  \qquad \,\,\,\,\,\,\,\,\,\,
    \sum_{ \sbrl \alpha \leftrightarrow \beta \sbrr} \equiv \sum_{ \substack{\{S_\alpha\},\{S_\beta\}\\ 
    S_{\beta,1} = S_{\alpha,1}}} \, .
    \end{aligned}
    \end{equation}
The free-energy 
of the system can be obtained by combining the free-energy shifts involved in the different processes, as explained in~\cite{Mezard2001,Rivoire2004}:
\begin{equation} \label{eq:free-energy}
F = \sum_{\alpha=1}^{N_\bigtriangleup} \Delta F_{\alpha}^{(s)}
- \sum_{\langle \alpha, \beta \rangle} \Delta F_{\alpha \leftrightarrow \beta}^{(l)}
= \frac{1}{2} \sum_{\langle \alpha, \beta \rangle} \left( 
\Delta F_{\alpha \to \beta}^{({\rm iter})} + \Delta F_{\beta \to \alpha}^{({\rm iter})}
\right) - \frac{1}{2} \sum_{\alpha=1}^{N_\bigtriangleup} \Delta F_{\alpha}^{(s)} \, ,
\end{equation}
where 
$\langle \alpha , \beta \rangle$ denotes the sum over the  $3 N_\bigtriangleup /2$ nearest-neighbors plaquettes on the graph (the last equality simply comes from the fact that $\Delta F_{\alpha}^{(s)} = \Delta F_{\alpha \to \beta}^{({\rm iter})} + \Delta F_{\alpha \leftrightarrow \beta}^{(l)}$ by construction). The average entropy of the system is then given by $\langle S \rangle = \beta (\langle E \rangle - F)$.
Analogously, the total average energy can be written as
\[
\langle E \rangle = \sum_{\alpha=1}^{N_\bigtriangleup} \left \langle e _\alpha^{(s)} \right \rangle 
- \sum_{\langle \alpha, \beta \rangle} \left \langle e_{\alpha \leftrightarrow \beta}^{(l)} \right \rangle \, .
\]
\end{widetext}

Equations~(\ref{eq:cavity}) can be written for arbitrary (large) RRGs and are expected to become exact in the thermodynamic limit.
On each triangle of the RRG one can define three cavity plaquettes by removing one of its three neighbors. Thus, Eqs.~(\ref{eq:cavity}) represent a set of $32 \times 3 \times N_\bigtriangleup$ coupled nonlinear algebraic equations for the $32$ marginal probabilities $p_{\alpha \to \beta}(\{S_\alpha\}|\{S_\beta\})$ associated to the $32$ possible configurations of the spins $\{S_\alpha\}$ on the cavity plaquette $\alpha$, given the configuration of the spins $\{S_\beta\}$. Once the fixed points of these equations is found, one can compute the marginal probabilities on each plaquette of the graph from Eq.~(\ref{eq:marginals}), along with the free-energy and all observables.
In the following, we will discuss three specific solutions of the equations in the thermodynamic limit, corresponding to the (RS) homogeneous paramagnet, the (RS) ordered crystalline state, and the (RSB) glassy phase.

\subsection{The paramagnetic phase}

The paramagnetic phase is characterized by translational invariance and corresponds to the homogeneous and RS solution of the recursion relations:
\[
p_{\alpha \to \beta}(\{S_\alpha\}|\{S_\beta\})  =  p_{\rm para}(\{S_\alpha\}|\{S_\beta\}) \, \qquad \forall \alpha,\beta \,  . 
\]
 The probabilities $p_{\rm para}(\{S_\alpha\}|\{S_\beta\})$ are given by the fixed point of Eqs.~(\ref{eq:cavity}) which in this limit become a simple system of $32$ coupled nonlinear algebraic equations. The free-energy, the energy, and the magnetization (which is identically zero by $Z_2$ inversion symmetry in the paramagnetic phase, which implies that $p_{\rm para}(\{S_\alpha\}|\{S_\beta\}) = p_{\rm para}(\{-S_\alpha\}|\{-S_\beta\})$), can be easily computed from Eqs.~(\ref{eq:marginals}),~(\ref{eq:obs}),~(\ref{eq:energy1}),~(\ref{eq:DFl}),~(\ref{eq:energy2}), and~(\ref{eq:free-energy}). 
This phase is expected to be stable at high temperature. However, 
the average entropy density $\langle s \rangle = \beta( \langle e \rangle - f)$ becomes  negative  below  a  certain temperature, $T_{s=0}(D)$.  This indicates that the homogeneous solution is certainly not appropriate to describe the low temperature region of the phase diagram.

\subsection{The stability of the paramagnetic phase}

The manifestation of the failure of the RS solution also shows up via a loss of stability of the RS fixed point, as given by a simple linear analysis.
To describe this instability one needs  to  introduce  a  probability distribution  ${\cal P} [ \vec{p} 
]$, where $\vec{p}$ is a short-hand notation for the $32$ marginal probabilities $p(\{S_\alpha\}|\{S_\beta\})$ and ${\cal P} [ \vec{p}]$ is defined as the probability density that the 
probabilities $p_{\alpha \to \beta}(\{S_\alpha\}|\{S_\beta\})$ on the cavity plaquette $\alpha$ 
are equal to $p(\{S_\alpha\}|\{S_\beta\})$.

In the homogeneous phase from Eq.~(\ref{eq:cavity}) one has that  the probability distributions of the marginal probabilities on the triangular plaquettes must satisfy the following self-consistent equation:
\begin{equation} \label{eq:stability}
{\cal P} [ \vec{p} ] = \int {\rm d} {\cal P} [ \vec{p}_\gamma ] \, {\rm d} {\cal P} [ \vec{p}_\delta ] \, \delta [ \vec{p} - \vec{\mathtt{p}} ( \vec{p}_\gamma, \vec{p}_\delta )] \, ,
\end{equation}
where $\vec{\mathtt{p}} ( \vec{p}_\gamma, \vec{p}_\delta )$ is a short-hand notation for the r.h.s. term of the recursion relations~(\ref{eq:cavity}). Close to the homogeneous paramagnetic solution we have, to first order,
\[
p (\{S_\alpha\}|\{S_\beta\})  \approx  p_{\rm para}(\{S_\alpha\}|\{S_\beta\}) + \delta p  (\{S_\alpha\}|\{S_\beta\}) \,  . 
\]
Starting with $\delta \vec{p}$ identically and independently distributed and injecting the expression above into Eq.~(\ref{eq:stability}), one has that the deviation of the marginal probabilities from the homogeneous solution evolves under iteration as
\[
\langle \delta \vec{p} \rangle = 2 \left .\frac{\partial \, \vec{\mathtt{p}} ( \vec{p}_\gamma, \vec{p}_\delta )}{\partial \vec{p}_\gamma} \right \vert_{\rm para} \!\! \langle \delta \vec{p} \rangle \, .
\]
where $\langle \cdot \rangle$ refers to the average using the distribution ${\cal P}(\vec{p})$. $\partial \, \vec{\mathtt{p}} / \partial \vec{p}_\gamma$ is actually a $32 \times 32$ Jacobian matrix. If $\lambda_{\rm max}$ denotes the eigenvalue of largest modulus of that matrix, the stability criterion simply reads $2 |\lambda_{\rm max}| \le 1$.
When $2 |\lambda_{\rm max}| > 1$, the paramagnetic solution is instead unstable with respect to a ``modulation'' instability, corresponding to a transition to a regime with successive (homogeneous) generations of the tree carrying different values of the marginal probabilities. Such modulation instability is thus a manifestation of an instability toward an ordered phase, which breaks translational invariance.

This instability criterion can also be obtained by studying response functions to a perturbation (which is related to correlations through the fluctuation-dissipation theorem)~\cite{Rivoire2004}. In this setting the instability is detected by means of the divergence of the linear magnetic susceptibility in the paramagnetic phase, defined as:
\[
\chi = \frac{1}{N_\bigtriangleup} \sum_{\alpha,\beta} \langle S_\alpha S_\beta \rangle_c = \frac{1}{N_\bigtriangleup} \sum_{\alpha,\beta} 
\left.
\frac{\partial \langle S_\alpha \rangle}{\partial h_\beta} \right|_{h_\gamma=0}\, ,
\]
where $S_\alpha \equiv \sum_{i \in \alpha} S_i$ is a short-hand notation for the magnetization of the plaquette $\alpha$ and $h_\beta$ is an external magnetic field conjugated to the magnetization of the plaquette $\beta$.
Making use of the homogeneity of the paramagnetic solution and the tree-like structure of the lattice, the susceptibility can be rewritten as
\[
\chi = 1 + 3 \sum_{r=1}^\infty 2^{r-1} \langle S_{\alpha_0} S_{\alpha_r} \rangle_c \, .
\]
where $S_{\alpha_0}$ and $S_{\alpha_r}$ are two plaquettes taken at distance $r$ on the tree. The series converges provided that $\lim_{r \to \infty} \log \langle S_{\alpha_0} S_{\alpha_r} \rangle_c/r < \log 2$. To evaluate $\langle S_\alpha S_\beta \rangle_c$, we invoke the fluctuation-dissipation relation:
\[
\langle S_{\alpha_0} S_{\alpha_r} \rangle_c = 
\left.
\frac{\partial \langle S_{\alpha_r} \rangle}{\partial h_{\alpha_0}} \right|_{h_\gamma=0}
\]
where $h_{\alpha_0}$ denotes the external magnetic field conjugate to $S_{\alpha_0}$. Since $h_{\alpha_0}$ is a function of (the components of) $P(\{S_{\alpha_0} \})$, we can use the chain rule along the branch of the tree which connects the plaquette $\alpha_0$ with the plaquette $\alpha_r$ through the plaquettes $\alpha_l$, $l=1, \ldots, r-1$:
\[
\frac{\partial \langle S_{\alpha_r} \rangle}{\partial h_{\alpha_0}} = \frac{\partial \langle S_{\alpha_r} \rangle}{\partial \vec{p}_{\alpha_r \to \alpha_{r-1}}} \left( \prod_{l=2}^{r} \frac{\partial \vec{p}_{\alpha_l \to \alpha_{l-1}}}{\partial \vec{p}_{\alpha_{l-1} \to \alpha_{l-2}}}\right) \frac{\partial \vec{p}_{\alpha_{1} \to \alpha_{0}}}{\partial \vec{P}_{\alpha_{0}}} \frac{\partial \vec{P}_{\alpha_{0}}}{\partial h_{\alpha_0}} \, .
\]
In the paramagnetic phase, all the intermediate marginal cavity probabilities are equal and the previous equation factorizes, leading again to $2 |\lambda_{\rm max}| \le 1$.

The maximal eigenvalue $\lambda_{\rm max}$ increases as the temperature is lowered and the linear susceptibility of the paramagnetic phase diverges at a certain temperature, signaling a modulation instability of the paramagnetic phase toward a crystalline phase (see Sec.~\ref{sec:crystal}) at a temperature $T_{\rm mod}(D)$.

One can also look for another kind of instability, namely a spin glass  instability, which manifests itself as a divergence of the non-linear susceptibility~\cite{Rivoire2004}, which is defined as
\[
\chi_{\rm sg} = \frac{1}{N_\bigtriangleup} \sum_{\alpha,\beta} \langle S_\alpha S_\beta \rangle_c^2 \, . 
\]
Equivalently, this instability appears as a widening of the variance $\langle ( \delta \vec{p} )^2 \rangle$ under the recursion of Eq.~(\ref{eq:stability}). Both approaches lead to a stability criterion $2 \lambda_{\rm max}^2 \le 1$. 
Note that this condition is always weaker than that for the modulation instability, $2 |\lambda_{\rm max}| \le 1$, associated to the crystalline order. However, it is the relevant one in the case of glassy phases, characterized by the establishment of long-range amorphous order.

Solving the recursion relations~(\ref{eq:cavity}) in the paramagnetic phase, we find that the homogeneous solution becomes unstable below a temperature $T_{\rm sg} (D)$, at which the spin glass susceptibility diverges (with $T_{s=0}(D) < T_{\rm sg}(D) < T_{\rm mod} (D)$).

This requires either a phase transition {\it before} the spin glass local instability is reached~\cite{krzakala2008,Biroli2002,Ciamarra2003,Weigt2003,Rivoire2004,Tarzia2007} (as occurs in the mean-field models of fragile glasses, described by a Random First-Order Transition~\cite{Kirkpatrick1989,Lubchenko2007}), or a continuous (possibly spin glass) transition {\it at} $T_{\rm sg}$. We will show below that the latter scenario is the correct one for the DKIAFM.
In order to do this 
in Sec.~\ref{sec:glass} we look for a solution of the recursion relations which breaks the replica symmetry, corresponding to a glassy phase where many local minima of the free-energy exist and where the local marginal probabilities fluctuate from a plaquette to another.

\begin{figure}
\includegraphics[width=0.38\textwidth]{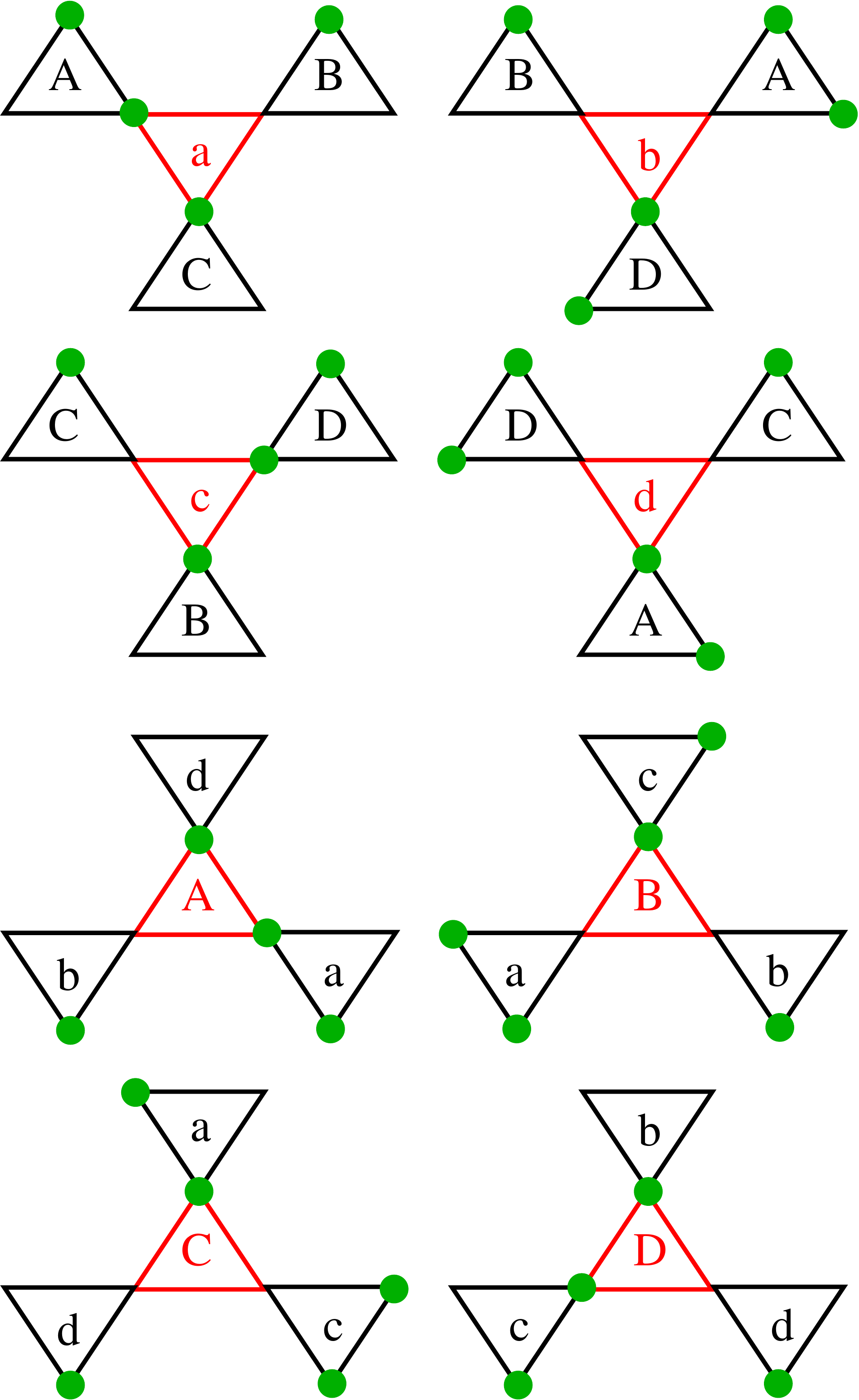}
\caption{\label{fig:crystal} Sketch of the sublattice structure introduced to describe the 
ordered crystalline phase proposed in~\cite{Chioar2016} and detected  in~\cite{Hamp2018}. The $4$ up-type plaquette sublattices are denoted by $A,B,C,D$, and the $4$ down-type plaquette sublattices are denoted by $a,b,c,d$. The green circles represent the $+1$ spins in one of the sixfold degenerate ground state configurations. Such sublattice structure is associated to $24$ different kind of cavity plaquettes, denoted by $A_n,B_n,C_n,D_n$, and $a_n,b_n,c_n,d_n$, with $n=1,2,3$, obtained by removing one of the three neighboring plaquette to each of the $8$ kind of sublattice plaquettes. For exemple, the cavity plaquette $a_1$ is obtained by removing the bottom neighboring plaquette $C$ from the plaquette $a$, and it is connected to the cavity plaquettes $A_3$ on the right and $B_2$ on the left, obtained respectively by removing the right neighboring plaquette $a$ from the plaquette of type $A$ and the left neighboring plaquette $a$ from the plaquette of type $B$.}
\end{figure}

\subsection{The crystal phase} \label{sec:crystal}

One can look for a crystalline RS solution, where the marginal probabilities do not fluctuate from  site  to  site,  but  are  different  in  different  sites (break-down of translational invariance). 
The (sixfold degenerate) crystalline state proposed in~\cite{Chioar2016} and observed numerically in~\cite{Hamp2018} is characterized by a $12$-spin unit cell and breaks (twofold) time-reversal symmetry and (threefold) rotation symmetry (see Refs.~\cite{Chioar2016,Hamp2018} for more details). In order to be able to account for such ordered phase we need to introduce $8$ sublattices of triangular plaquettes (see Fig.~\ref{fig:crystal}), corresponding to $24$ different sets of cavity plaquettes. 
The  merging of the $24$ cavity plaquettes is  done  taking  into  account  the structure  of  the  crystalline  phase, as explained in the caption of Fig.~\ref{fig:crystal}. The recursion equations~(\ref{eq:cavity}) become then a set of $24 \times 32$ coupled nonlinear algebraic equations for the marginal probabilities on the $24$ cavity plaquettes on each sublattice. The solution of these equations appears  discontinuously at  a spinodal  point $T_{\rm sp}(D)$, and becomes  thermodynamically  stable when the corresponding free-energy crosses the paramagnetic one, at the melting temperature $T_m(D)$. At that temperature we observe a first-order phase transition characterized by a spontaneous breakdown of the translational, rotational, and spin inversion invariance, accompanied by a discontinuous  jump  of  the energy density and of the entropy density. Decreasing further the temperature, the energy in the crystalline  phase  approaches  quickly  the ground state value, $e_{\rm GS} \approx -1.6116$ (which turns out to be remarkably close to the one found with Monte Carlo simulations of systems of $300$ spins, $e_{\rm GS} \approx -1.515$ ~\cite{Hamp2018}), and  the  entropy quickly approaches zero.

Inspecting the (ground state) spin configuration of Fig.~\ref{fig:crystal}, it was noticed in~\cite{Hamp2018} that one of the three spins of the Kagome triangles are completely polarized (i.e., the bottom spins of sublattices $a,b,c,d$ and the top spins of sublatices $A,B,C,D$), 
with the state having zero magnetization overall. 
Note that the need to introduce $8$ sublattices of triangular plaquettes is due to the fact that the spin pattern on the two non-polarized rows of spins of the Kagome triangles (i.e., along the horizontal bonds in Fig.~\ref{fig:crystal}) has period four, with three spins $S = \mp 1$ followed by one spin $S=\pm 1$.
Based on these observations, suitable order parameters for the transition to the ordered state are the sublattice magnetizations:
\[
m_X = \sum_{i \in X} S_i \, ,
\]
with $X$ denoting the $8$ different sublattices: $X=\{A,B,C,D,a,b,c,d\}$. 
These order parameters essentially coincide with the emergent effective charge variables introduced in Ref.~\cite{Hamp2018}, derived from the so-called dumbbell picture~\cite{Castelnovo2008}.  The (ground state) spin configuration of Fig.~\ref{fig:crystal} corresponds to $m_a=m_c=m_A=m_D= \pm 1$ and $m_b=m_d=m_B=m_C= \mp 1$. 
Equivalently, one can choose as order parameter $m_{\rm pol} =  \langle S_{\rm pol} \rangle $, the average magnetization of the spins of the Kagome triangles that are completely polarized, as done in~\cite{Hamp2018}.
Following this suggestion, we will use $|m_{\rm pol}|$ as the order parameter jumping from zero to a finite value at the transition.

We have also looked for other plausible competing ordered phases, which break the translational and rotational symmetries in different ways and have a different unit cells. 
However, such alternative crystalline states turn out to be less favourable (i.e., they have a higher free-energy) compared to the crystalline phase of~\cite{Chioar2016,Hamp2018}. Yet, if one does not include the dipolar interactions between the second nearest-neighboring plaquettes, the crystalline phase depicted in Fig.~\ref{fig:crystal} disappears (i.e., no physically relevant fixed point of the recursion relations is found corresponding to the sublattice structure of Fig.~\ref{fig:crystal}), and another completely different (fourfold degenerate) ordered phase emerges.
This observation highlights the importance of accounting for the dipolar interaction as accurately as possible, in order to recover the correct description of the ordered phase~\cite{cutoff}.

\subsection{The spin glass phase} \label{sec:glass}

The  paramagnetic  phase  is  metastable  below $T_m (D)$, corresponding to a supercooled regime. However, as mentioned 
above, the predicted entropy density becomes negative as the temperature is lowered below $T_{s=0}(D)$, implying that this solution does not describe well the low temperature region. 
Moreover, the homogeneous solutions becomes unstable below a certain temperature $T_{\rm sg} (D)>T_{s=0}(D)$, at which the spin glass susceptibility diverges. 
The ``entropy crisis'' and the spin glass instability are manifestations of the appearence of a huge number of metastable glassy states. The RS approach  fails  because  it  does  not  take  into  account  the  existence of several local minima of the free-energy.
This requires either a phase transition before the spin glass local instability is reached (as in the case of lattice models for fragile glasses in the mean-field limit~\cite{krzakala2008,Biroli2002,Ciamarra2003,Weigt2003,Rivoire2004,Tarzia2007} described by a Random First-Order Transition~\cite{Kirkpatrick1989,Lubchenko2007}), or a continuous spin glass transition at $T_{\rm sg}$.
In order to understand which of these two possible scenarios is the correct one for the DKIAFM, we have to look for a solution of the recursion relations which breaks the replica symmetry, corresponding to a glassy phase where many local minima of the free-energy exist and where the local marginal probabilities fluctuate from a plaquette to another.
We thus need to perform a statistical treatment of sets of solutions of Eq.~(\ref{eq:cavity}). The simplest setting which allows to proceed further in this direction is provided by a one-step RSB ansatz, which starts from the assumption that exponentially many (in $N_\bigtriangleup$) solutions of the recursion relations exist. 
More precisely, we assume that the number ${\cal N} (f)$ of solutions with a given free-energy density $f$ on graphs of size $N_\bigtriangleup$ is ${\cal N} (f) \sim \exp [ N_\bigtriangleup \Sigma (f)]$, where $\Sigma(f) \ge 0$ is called the {\it configurational entropy} (or {\it complexity}) and is supposed to be an increasing and concave function of the free-energy $f$. This is a strong hypothesis which is justified by its self-consistency. Under these assumptions, one can show that the 1RSB self-consistent equation for the probability distribution of the marginal cavity probabilities becomes~\cite{Mezard2001,Rivoire2004}
\begin{equation} \label{eq:1RSB}
{\cal P}_m [ \vec{p} ] \propto \int {\rm d} {\cal P}_m [ \vec{p}_\gamma ] \, {\rm d} {\cal P}_m [ \vec{p}_\delta ] \, \delta [ \vec{p} - \vec{\mathtt{p}} ( \vec{p}_\gamma, \vec{p}_\delta )]
\, e^{-\beta m \Delta F^{({\rm iter})}} \, ,
\end{equation}
where $\vec{\mathtt{p}} ( \vec{p}_\gamma, \vec{p}_\delta )$ is a short-hand notation for the r.h.s. term of the recursion relations~(\ref{eq:cavity}) and $\Delta F^{({\rm iter})}$ is the free-energy shift involved in the iteration process defined in Eq.~(\ref{eq:cavity}) via the normalization of the cavity marginal probabilities.
The probability distribution depends on the parameter $m$ which is the  breakpoint  in  Parisi's order parameter function at the 1RSB level~\cite{Mezard1987,Mezard2001,Rivoire2004}, and is defined as $m = (1/\beta) \partial \Sigma / \partial f$ (all the details of the calculation can be found in Refs.~\cite{Mezard2001,Rivoire2004,krzakala2008}).
Similarly to Eq.~(\ref{eq:free-energy}), the 1RSB free-energy density functional is given by
\begin{equation} \label{eq:free-1RSB}
\phi(m) = \Delta \phi^{(s)}(m) - \frac{3}{2} \Delta \phi^{(l)} (m) \, ,
\end{equation}
with
\[
\begin{aligned}
e^{-\beta m \Delta \phi^{(s)}} & = \int {\rm d} {\cal P}_m [ \vec{p}_\beta ] \, {\rm d} {\cal P}_m [ \vec{p}_\gamma ] \, {\rm d} {\cal P}_m [ \vec{p}_\delta ] \, e^{-\beta m \Delta F^{(s)}} \, , \\
e^{-\beta m \Delta \phi^{(l)}} & = \int {\rm d} {\cal P}_m [ \vec{p}_\alpha ] \, {\rm d} {\cal P}_m [ \vec{p}_\beta ] \, e^{-\beta m \Delta F^{(l)}} \, ,
\end{aligned}
\]
where the free-energy shifts have been defined in Sec.~\ref{sec:cavity}.
The other relevant thermodynamic observables, such as, e.g., the average energy, can be obtained in a similar fashion~\cite{Mezard2001,Rivoire2004}.
The parameter $m$ is fixed by the maximization of the free-energy functional with respect to it~\cite{Mezard2001,Rivoire2004}, which allows to recover the complexity as a Legendre transform of $\phi(m)$:
\[
m \, \phi (m) = m f - \frac{1}{\beta} \Sigma (f) \, .
\]
The RS high-temperature homogeneous description of the phase is recovered by taking 
$P_m(\vec{p}) = \delta (\vec{p} - \vec{p}_{\rm para})$ and $m=1$~\cite{remark2}. 

Since Eq.~(\ref{eq:1RSB}) is a functional relation, an analytical treatment is not possible in general. Yet the self-consistent equation can be efficiently solved numerically with arbitrary precision using a population dynamics algorithm (for all technical details see~\cite{Mezard2001}). For high values of the temperature ($T>T_{\rm sg} (D)$) we recover the paramagnetic solution. Lowering the temperature, a nontrivial solution of the 1RSB equation   appears continuously exactly at $T_{\rm sg} (D)$. Right below $T_{\rm sg}$ the probability distribution ${\cal P}_m (\vec{p})$ acquires an infinitesimal widening of the variance $\langle (\delta \vec{p})^2 \rangle$. 
This scenario corresponds to a continuous transition to a spin glass phase at the temperature at which the spin glass susceptibility diverges. 
The order parameter of the spin glass transition is the Edwards-Anderson order parameter, $q_{\rm EA} = (1/N) \sum_i \langle S_i \rangle^2$, which vanishes linearly as $q_{\rm EA} \sim (T_{\rm sg} - T)$ for $T \to T_{\rm sg}^-$~\cite{Mezard1987}.

As it is well-known, the low-temperature spin glass 
phase should be described by full RSB~\cite{Mezard1987}. However, any new level of RSB will require considering a more sophisticated situation, namely a distribution over the probability distribution of the previous level. For instance, the two-step RSB will be written as a distribution ${\cal Q}[{\cal P}[\vec{p}]]$ over distributions ${\cal P}(\vec{p})$. Describing with this formalism a finite connectivity system with full RSB is therefore too complicated, and we will limit ourselves to the 1RSB Ansatz. 
Moreover, since solving the  self-consistent functional equation~(\ref{eq:1RSB}) via  population  dynamics  is  quite computationally demanding, we did not perform the maximization of the free-energy functional~(\ref{eq:free-1RSB}) with respect to $m$. For these reasons, our approach only provides an approximate description of the equilibrium properties of the spin glass phase and we have not pushed the 1RSB calculations far below $T_{\rm sg}$ (essentially we only consider few values of the temperature in the vicinity of the critical point).

\begin{figure}
\includegraphics[width=0.5\textwidth]{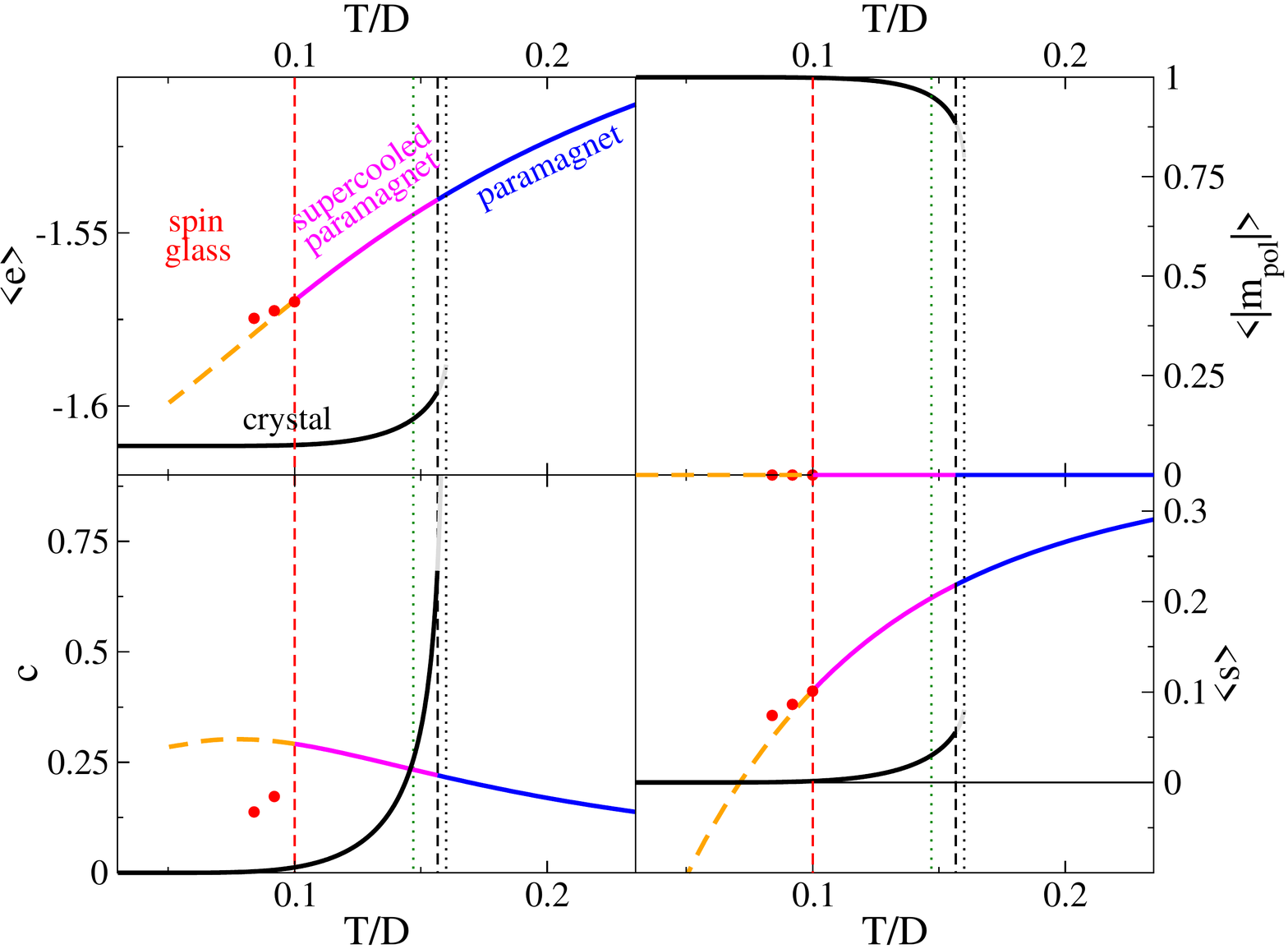}
\caption{\label{fig:observables} Average energy density $\langle e \rangle$ (top left),  magnetization of the polarized spins $|m_{\rm pol}|$ (top right), specific heat (per plaquette) $c=\partial \langle e \rangle / \partial T$ (bottom left), and average entropy density $\langle s \rangle$ (bottom right) as a function of the temperature $T$ for $J=0.5$ and $D=1$. Data in the paramagnetic phase are shown in blue, in the crystal phase in black,  in the  supercooled paramagnetic phase in magenta. The red circles are obtained by solving the 1RSB equations in the spin glass phase (with $m=1$). The vertical dashed black and red lines correspond to the first-order transition to the crystal state at $T_m$ and the continuous transition to the spin glass phase at $T_{\rm sg}$, respectively. The gray curves correspond to the crystal solution in the metastable region and end at the spinodal point ($T_{\rm sp}$, vertical black dotted line). The green dotted vertical line gives the position of the modulation instability of the homogeneous solution, $T_{\rm mod}$. The orange dashed curves correspond to the (unstable) RS solution of the equation below the spin glass transition point, and show the entropy crisis \`a la Kauzmann of the RS solution (at $T_{s=0}$).}
\end{figure}

\section{Phase diagram and thermodynamic behavior} \label{sec:results}

In this Section we discuss the main results found within the mean-field treatment of the DKIAFM described in the previous sections. In order to compare with the numerical results of the Monte Carlo simulations of~\cite{Hamp2018}, we start by fixing the parameter $D$ to $1$, as in Refs.~\cite{Hamp2018,Chioar2016}, and measure several observables such as the average energy density $\langle e \rangle = \langle E \rangle/N_\bigtriangleup$, the (intensive) specific heat $c = \partial \langle e \rangle / \partial T$, the magnetization of the polarized spin in one of the sixfold degenerate ground state configurations $|m_{\rm pol}|$, and the average entropy density $\langle s \rangle = \langle S \rangle/N_\bigtriangleup$, as a function of the temperature $T$ in the paramagnetic, crystal, and 1RSB glass solutions of the recursive cavity equations. The results are shown in Fig.~\ref{fig:observables}.
At high temperature the system is found in the paramagnetic phase. Upon lowering the temperature, a first-order transition to the crystalline phase proposed in Refs.~\cite{Hamp2018,Chioar2016} (see Fig.~\ref{fig:crystal}) occurs at $T_m$. The order parameter $|m_{\rm pol}|$ presents a finite jump at $T_m$, where the average energy and entropy densities also display an abrupt decrease.
The transition to the ordered state turns out to be very weakly first-order, in the sense that the spinodal point of the crystalline solution, $T_{\rm sp} \approx 0.16 \, {\rm K}$, is very close to the transition temperature $T_m \approx 0.1566 \, {\rm K}$ where the free-energies of the paramagnetic phase and the crystal phase cross. These temperatures are also numerically close to the modulation instability temperature $T_{\rm mod} \approx 0.147 \, {\rm K}$. Since the specific heat of the crystal solution diverges at the spinodal point, the vicinity of $T_m$ and $T_{\rm sp}$ results in a very large jump (of about a factor $3$) of the intensive specific heat at the transition. This feature might explain the deviations observed in the numerical simulations of the expected scaling of the peak of the (extensive) specific heat as $C_{\rm max} \propto N_\bigtriangleup$~\cite{Hamp2018}.

Although approximate, our approach accounts remarkably well for the numerical results of Ref.~\cite{Hamp2018}. As expected, the transition temperature is overestimated by the mean-field approximation (by about a factor $3$). Yet, the temperature dependencies of the specific heat, the energy, and the magnetization are, also at a quantitative level, very similar to the ones found in Ref.~\cite{Hamp2018} (recall that the energy, entropy, and specific heat {\it per spin} are obtained by multiplying the energy, entropy, and specific heat {\it per plaquette} by a factor $2/3$).

The  paramagnetic  phase  is  metastable  below $T_m$, corresponding to a supercooled regime.
If one keeps lowering the temperature within the supercooled phase, the spin glass susceptibility grows and diverges at $T_{\rm sg} \approx 0.1 \, {\rm K}$, where a continuous transition to a spin glass phase takes place. Although the spin glass phase is presumably described by full RSB (at least at the mean-field level), our approach only allows one to perform an approximate 1RSB Ansatz for the low temperature glassy phase. Moreover, solving the self-consistent functional equation~(\ref{eq:1RSB}) via population dynamics is computationally heavy. For these reasons, we did not push the calculations of the thermodynamic observables too deep into the spin glass phase, and only solved the equations for few points close to the critical temperature.

\begin{figure}
\includegraphics[width=0.49\textwidth]{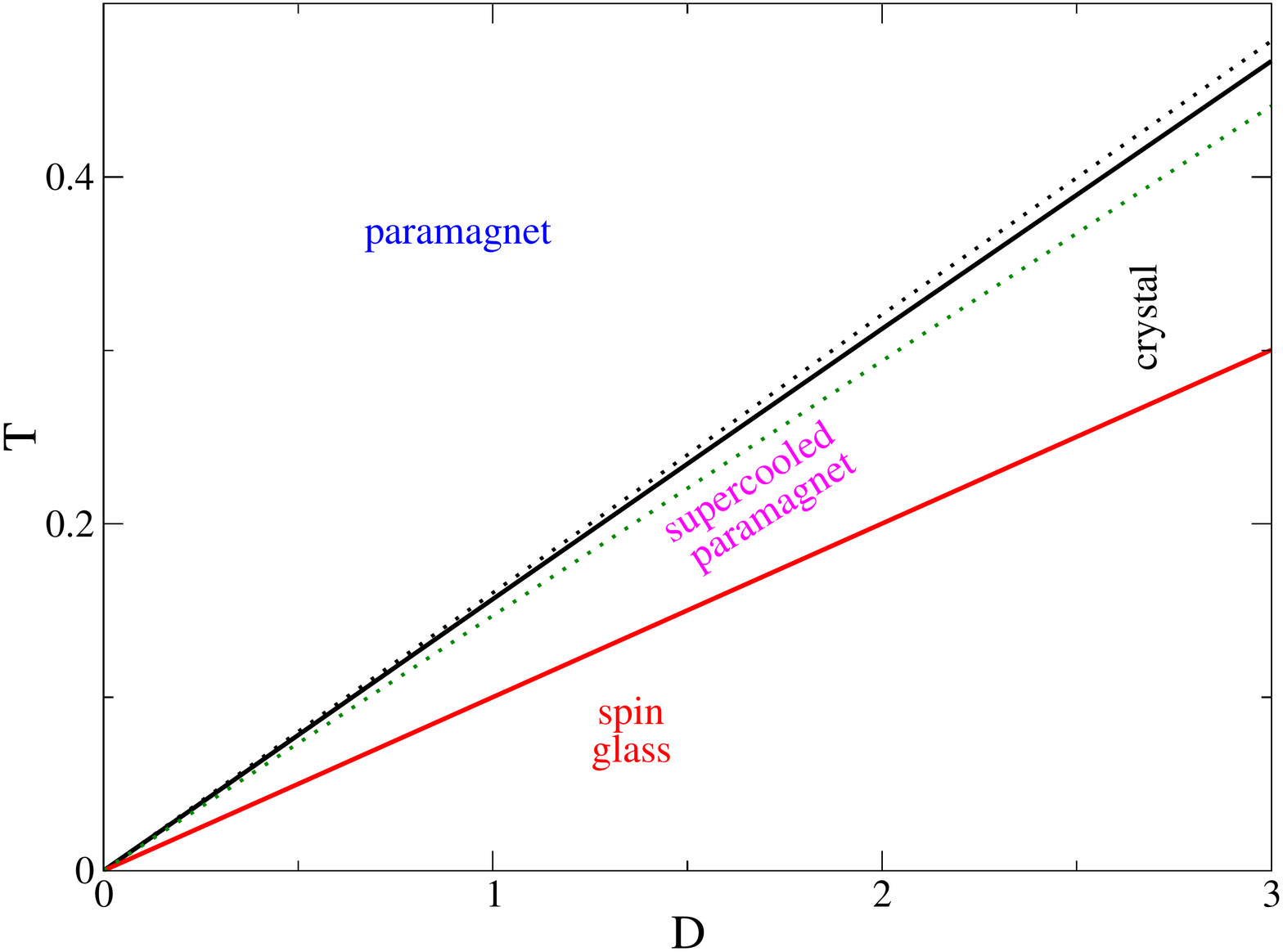}
\caption{\label{fig:pd} Mean-field phase diagram of the model in the $D$-$T$ plane for $J=0.5$, showing the position of the different phases and the transition lines. The black continuous line corresponds to the first-order melting transition, $T_m(D)$, where the free-energies of the paramagnetic and the crystalline solutions cross. The black dotted line gives the spinodal point at which the crystalline solution appears discontinuously. The green dotted line corresponds to the modulation instability of the paramagentic solution, $T_{\rm mod} (D)$. The red continuous line is the continuous spin glass transition, $T_{\rm sg} (D)$, where the spin glass susceptibility diverges. }
\end{figure}

In Fig.~\ref{fig:pd} we plot the phase diagram of the DKIAFM, showing the position of the different phases when varying the temperature $T$ and the dipolar interaction $D$ ($J$ is fixed to $J=0.5 \, {\rm K}$).
The effect of varying the dipolar interaction turns out to be particularly simple. In fact, we find that the phase boundaries, as well as all the characteristic temperature scales, vary linearly with $D$:
\[
\begin{aligned}
T_m & \approx 0.1566 \, D \, ,\\
T_{\rm sg} & \approx 0.1 \, D \, , \\
T_{\rm sp} & \approx 0.16 \, D \, , \\
T_{\rm mod} & \approx 0.147 \, D \, , \\
T_{\rm s=0} & \approx 0.0713 \, D \, .
\end{aligned}
\]
As expected, in the limit $D=0$ the paramagnetic phase is stable at all temperatures and corresponds to the only solution of the recursion relations. This is due to the fact that for $D=0$ the system is much less frustrated and has a highly (i.e., extensively) degenerate ground state (i.e., $\langle s \rangle$ approaches a finite value in the $T \to 0$ limit), since each plaquette has a sixfold degenerate ground state which corresponds to the ice rule (two $+1$ and one $-1$ spins or two $-1$ and one $+1$ spins). In particular, for $D=0$ the model reduces to the nearest-neighbor Kagome spin ice model of Wills, Ballou, and Lacroix~\cite{Wills2002}, for which a Pauling estimate yields the entropy $s_{\rm GS} = (3/2) \log[2 (3/4)^{2/3}] \approx 0.75225$, while our mean-field approximation yields $s_{\rm GS} \approx 0.75204$. 
When the dipolar interactions are turned on ($D>0$), such degeneracy is lifted, and a specific crystalline ground state structure emerges. The minimization of the local interactions produces a much stronger geometric frustration for $T \lesssim D$ and gives rise to the emergence of a spin glass phase at low temperatures, characterized by an extremely rough free-energy landscape (at least at the mean-field level).
The fact that all the relevant temperature scales of the problem show an apparent linear dependence on $D$ is precisely due to the fact that the relevant energy scale is the energy difference between the ground state and the first excited states, which goes linearly to zero with $D$.

\section{Conclusions} \label{sec:conclusions}

In this paper we have developed an analytical mean-field treatment for the equilibrium properties of the DKIAFM introduced in~\cite{Chioar2016} and studied numerically in~\cite{Hamp2018}. 
Our mean-field approach is based  on  a  cluster  variational Bethe-Peierls formalism~\cite{Jaubert2008,Foini2013,Levis2013,Cugliandolo2015,Pelizzola2005} and on the cavity method~\cite{Mezard2001}, and consists in studying the model on a sparse random tree-like graph of triangular Kagome plaquettes, and cutting-off the dipolar interaction beyond the second nearest-neighbor plaquettes (i.e., the $5$th nearest-neighbor spins).
Our results essentially confirm and support the observations reported in Ref.~\cite{Hamp2018},
which were obtained using Monte Carlo simulations of relatively small system (ranging from $48$ to $300$ spins), and  might  be  affected  by  both  strong  finite-size  effects
and by the difficulty of reaching thermal equilibrium in a reliable fashion due to strong metastability effects.

The summary of our results is the following.
Upon decreasing the temperature we first find a transition to a sixfold degenerate crystal state which breaks time reversal, translation,  and  rotation  symmetry  as  the
one proposed in~\cite{Chioar2016,Hamp2018}. Such transition is indeed discontinuous, as suggested in~\cite{Hamp2018}, although its first-order character turns out to be extremely weak, which might explain the strong finite-size effects observed in the finite-size scaling of the numerical data of the specific heat.
When the system is supercooled below the first-order
transition, we find that the paramagnetic state becomes
unstable below a temperature at which the spin glass susceptibility diverges and a continuous spin glass transition takes place at the mean-field level. 

On the one hand, the results presented here support and clarify
the numerical findings of Ref.~\cite{Hamp2018}. On the other hand, they provide a first step to  bridge  the  gap  between  the  slow  dynamics  observed in geometrically frustrated magnetic systems and the mean-field theory of glassy systems formulated in terms of
rough free-energy landscape.  We believe that this analysis is of particular interest, especially in the light of the experimental relevance of the model, which could be potentially  realized  in  several  realistic  set  ups,  including
colloidal crystals~\cite{Han2008,Zhou2017,Tierno1,Tierno2,Tierno3,Colloidal}, artificial nanomagnetic arrays~\cite{Chioar2014,Nisoli2013,Heyderman1,Luning}, polar molecules~\cite{Ni2008}, atomic gases with large magnetic dipole moments~\cite{Griesmaier2005}, and layered bulk Kagome materials~\cite{Scheie2016,Paddison2016,Dun2017}.

Some comments are now in order.

The lower-critical dimension of the spin glass transition 
is expected  to  be  
$d_L \approx 2.5$~\cite{Franz1994} (at  least  in  the  case
of  short-range  interactions).  Hence  we do not expect a genuine spin glass phase for the DKIAFM. Yet in $2d$ the spin glass amorphous order can establish over very large (although not infinite) length scales and the spin glass susceptibility can become very large
(although  not  infinite)  at  low  temperature  due  to  the
vestiges of the transition.  Indeed, there are plenty of experimental studies using thin films that at sufficiently low temperatures 
behave as the $3d$ counterparts. See, e.g.~\cite{Gucchhait2017} for a 
very recent reference and~\cite{Mattson1992}  
for a more classical ones. The situation is similar concerning numerical simulations~\cite{Fernandez2019}.

Concerning the dynamics, very early Monte Carlo simulations of the $3d$ Edwards-Anderson model suggested that the spin auto-correlation function, close but above the expected critical temperature, decays as a stretched exponential~\cite{Ogielski1985}.
Therefore, although the model is expected to have a conventional second order phase transition with critical slowing down and algebraic decay of correlation functions, 
for the system sizes and time-scales 
accessed in this paper, the time-delayed correlations 
were satisfactorily fitted by such an anomalous
form, with a stretching exponent decaying with decreasing temperature.
Just a bit later, in~\cite{OgielskiMorgenstern1985} the conventional critical slowing down was recovered. 
A stretched exponential relaxation
of the self-correlation in the DKIAFM was reported in~\cite{Hamp2018}.
However, our results suggest that at sufficiently large time and length scales this behavior might be replaced by a conventional power law decay for control parameters in the critical region.

The case $J=0$ and $D>0$~\cite{Takagi1993} has been left over by the present investigation and might be an interesting subject for future studies. Possibly, the most interesting questions would be the  investigation of  how the properties of the model are  affected  by  quantum  fluctuations.

\begin{acknowledgments}
We would like to thank C. Castelnovo for enlightening and helpful discussions.
L. F. Cugliandolo and M. Tarzia are members of the {\it Institut Universitaire de France}.
This work is supported by ``Investissements d'Avenir" LabEx PALM (ANR-10-LABX-0039-PALM) (EquiDystant project, L. Foini).
\end{acknowledgments}


\begin{thebibliography}{99}
\bibitem{remark} Note that the fact that the mean-field value of the ground state energy is found to be lower than the $2d$ value is due to the fact that the dipolar interactions are cut-off beyond the  second  nearest-neighbour plaquettes in the mean-field calculations.
\bibitem{remark2} In particular, it can be shown that $m= 1$ always gives back the paramagnetic RS solution.
\end{thebibliography}
\end{document}